\documentclass[letterpaper, 11 pt]{article}  

\usepackage[dvips]{graphicx}  
\usepackage[latin1]{inputenc} 
\usepackage{amssymb}          
\usepackage{amsmath}          
\usepackage{amsfonts}
\usepackage[usenames,dvipsnames]{xcolor}
\usepackage{curves}
\usepackage{comment}
\usepackage{a4wide}

\newtheorem{lemma}{Lemma}

\newtheorem{prop}{Proposition}

\newcommand{\ket}[1]{|#1\rangle}
\newcommand{\bra}[1]{\langle #1|}
\newcommand{\braket}[2]{\langle #1|#2\rangle}

\newcommand{\Hi}{\mathcal{H}}

\newcommand{\trace}{\mathrm{{Tr}}}

\newcommand{\q}[1]{\vert #1 \rangle}
\newcommand{\qd}[1]{\langle #1 \vert}

\newcommand{\pope}{\mathsf{p}}

\newcommand{\Ec}{\mathcal{E}}

\newcommand{\N}{\mathcal{N}}

\newcommand{\qed}{\hfill $\Box$ \vskip 2ex}

 \newcommand{\beq}{\begin{equation}}
 \newcommand{\eeq}{\end{equation}}
 \newcommand{\beqa}{\begin{eqnarray}}
 \newcommand{\eeqa}{\end{eqnarray}}
 \newcommand{\beqan}{\begin{eqnarray*}}
 \newcommand{\eeqan}{\end{eqnarray*}}
 \newcommand{\bea}{\begin{eqnarray}}
 \newcommand{\eea}{\end{eqnarray}}
 
\newcommand{\ft}{\color{black}}
\newcommand{\ftl}{\color{black}}

\newcommand{\as}{\color{black}}

\newcommand{\noise}{Lindblad } 
\newcommand{\proof}{\emph{Proof:~}} 

\title{\LARGE \bf
Symmetrization for Quantum Networks:\\ a Continuous-Time Approach \thanks{Work partially supported by the University of Padua QFUTURE grant and by the Belgian Network DYSCO (Dynamical Systems, Control, and Optimization), funded by the Interuniversity Attraction Poles Programme initiated by the Belgian Science Policy Office.}
}

\author{Francesco Ticozzi\thanks{F. Ticozzi is with Dipartimento di Ingegneria
dell'Informazione, Universit\`a di Padova, via Gradenigo 6/B,
35131 Padova, Italy ({\tt ticozzi@dei.unipd.it}) and Dept. of Physics and Astronomy, Dartmouth College, 6127 Wilder, 03755 Hanover, NH (USA).}%
, Luca Mazzarella\thanks{L. Mazzarella is with Dipartimento di Ingegneria
dell'Informazione, Universit\`a di Padova, via Gradenigo 6/B,
35131 Padova, Italy ({\tt mazzarella@dei.unipd.it}).}
 and Alain Sarlette\thanks{A. Sarlette is with SYSTeMS, Ghent University, Technologiepark Zwijnaarde 914, 9052 Gent, Belgium ({\tt alain.sarlette@ugent.be}).}
}

\begin{document}

\maketitle
\thispagestyle{empty}
\pagestyle{empty}

\begin{abstract}
In this paper we propose a continuous-time, dissipative Markov dynamics that asymptotically drives a network of $n$-dimensional quantum systems to the set of states that are invariant under the action of the subsystem permutation group. The Lindblad-type generator of the dynamics is built with two-body subsystem swap operators, thus satisfying locality constraints, {\ft and preserve symmetric observables.} The potential use of the proposed generator in combination with local control and measurement actions is illustrated with two applications: the generation of a global pure state and the estimation of the network size.
\end{abstract}

\section{INTRODUCTION}

Classical consensus algorithms and the related distributed control problems have recently generated an impressive body of literature, motivated by applications in distributed computation and multi-agent coordination, see e.g.~\cite{Gorinevsky2008,Bamieh2002a,D'Andrea2003,BulloBook,BarHesp1,VaraiyaEst82,BoydADMM,leonard2007collective,ZampieriGen}.
We have recently recast the problem, and a class of algorithms for its solution, as a dynamical symmetrization problem in a group-theoretic framework \cite{mazzarella-symmetrization}. This allows to extend the use of simple and robust algorithms, e.g.~of gossip type \cite{BoydGossip}, to new settings and applications. Among these, we have studied a quantum version of consensus problems and its applications \cite{mazzarella-TAC}, {\as which can be seen as symmetrization with respect to the subsystem-permutation group, {\ft as well as control methods like quantum dynamical decoupling \cite{viola-dd}, in which we generally do not have a multipartite structure and symmetrization is attained with respect to other finite groups}}.
%
%
The emerging dynamics are intrinsically in discrete time, and suitable for sequential implementation in dissipative quantum simulators \cite{barreiro}.

In this work, we extend the same line of research to {\em continuous-time} dynamics bound to satisfy some locality constraint. This is done in the same spirit as \cite{ticozzi-QL,ticozzi-QIP}, and connects with the idea of dissipative quantum computation in continuous time \cite{Verstraete2009}.

Enforcing symmetrization using continuous quantum dynamical semigroups requires different tools and 
{\as provides specific results that complement the discrete-time results. So, while the general symmetrization viewpoint of \cite{mazzarella-symmetrization} still applies, the formulation/implementation and evolution of the associated dynamics brings up some practically relevant novelties.}
A particularly attractive feature is that a continuous-time algorithm can implement several driving influences simultaneously on a single subsystem, which will allow us to apply diverse symmetrizing actions in some sense ``in parallel''; {\as in the discrete-time case, {\ft on the other hand}, a single {\ft quasi-local} action must be selected at each step (therefore the gossip type approach).} This could lead to an advantage {\as in convergence time} and additional robustness for settings in which {\as continuous-time implementations with many parallel actions} could be viable.
{\as This property is in agreement with our quest for a \emph{robust}, flexible way of \emph{asymptotically} controlling a quantum network towards a target state. {\ft However, when working with accurately-engineered dynamics} in a well-known network, discrete-time operation in principle offers the possibility of designing algorithmic procedures with finite-time convergence {while in the continuous setting the convergence is necessarily asymptotic}.}
The same type of difference emerges in the effort of engineering entangled states on quantum networks \cite{ticozzi-inprep}.

In this paper we propose a class of continuous-time Quantum Dynamical Semigroup (QDS) generators that achieve symmetrization under strict locality constraints. We prove the convergence of the dynamics and completely characterize its asymptotic behavior. {\as For this conference session we restrict our scope to symmetrization with respect to subsystem permutations in a network.} Our promising results in this direction should suggest how to transpose more general settings \cite{mazzarella-symmetrization} to continuous-time operations, and stimulate the search for new compelling applications for continuous symmetrizing dynamics.

The paper is organized as follows. Section II introduces the reader to the model for the quantum network, the dynamics and the locality notion. The symmetrizing dynamics is introduced in Section III, where we also prove its convergence properties. Lastly, some control motivated applications are proposed in Section IV.



\section{PRELIMINARIES}


\subsection{Multipartite Quantum Systems and their Symmetric States}\label{sec:system}

In quantum theory, each locally-accessible quantum system is associated to a Hilbert space, and a representation for a global system containing more than one subsystem is obtained by constructing a (larger) Hilbert space which is the tensor product of those corresponding to its constituents\footnote{{\ft In the following sections we shall use Dirac's notation  \cite{sakurai}, as it customary in the quantum control literature: $\ket{\psi}$ will denote an element of $\Hi$ and $\bra{\psi}$ a dual vector, i.e. linear functional on $\Hi$ }}.
{\as For any Hilbert space $\Hi$,} let ${\mathfrak B}(\Hi)$ represent the set of linear operators on $\Hi,$ 
${\mathfrak U}(\Hi)$ the unitary ones and ${\mathfrak H}(\Hi)$ the self-adjoint ones; {\as the latter characterize physical observables}. 

In this paper, we consider a multipartite system composed of $m$ identical finite-dimensional subsystems, labeled with indices $i=1,\ldots,m$ and with individual Hilbert spaces $\Hi_i \simeq \Hi$, $\dim(\Hi_i)=\dim(\Hi)=n\geqslant 2$. This multipartite system will represent our {\em quantum network}, with associated Hilbert space $\Hi^{\otimes m}:=\Hi_1\otimes \dots \otimes \Hi_m$. 
{\as A state of the network is characterized by a density operator on $\Hi^{\otimes m},$ namely the state space is the set ${\mathfrak D}(\Hi^{\otimes m})=\{\rho\in {\mathfrak H}(\Hi^{\otimes m})|\rho \text{ is positive semidefinite and }\;\trace(\rho)=1.\}.$}

For any operator $X \in \mathfrak{B}(\mathcal{H})$, we will denote by $X^{\otimes m}$ the tensor product $X \otimes X \otimes ... \otimes X \in \mathfrak{B}(\Hi^{\otimes m})$ with $m$ factors, and by $X^{(i)}\in \mathfrak{B}(\Hi^{\otimes m})$ the operator $X$ acting only on the $i$-th subsystem:
\[X^{(i)}:=I^{\otimes (i-1)}\otimes X \otimes I^{\otimes (m-i)}.\]
Operators of this form will be referred to as strictly {\em local} operators.

An important class of operators, which plays a central role in our work, is the class of operators which are invariant with respect to all subsystem permutations. Let us denote the set of all permutations of the first $m$ integers by ${\mathfrak P}_m,$ and $\rm e\in {\mathfrak P}_m$ be the trivial one.
Permutations of quantum subsystems are represented by unitary operators $U_{\pi} \in {\mathfrak U}(\Hi^{\otimes m})$, $\pi\in{\mathfrak P}_m$, which are uniquely defined by 
\[U_{\pi} (X_1\otimes\ldots \otimes X_m) U_{\pi}^\dag= X_{\pi(1)}\otimes\ldots \otimes X_{\pi(m)}\]
for any operators $X_1,\ldots X_m$ in ${\mathfrak B}(\Hi)$. 
{\as A state $\rho \in {\mathfrak D}(\Hi^{\otimes m})$ or observable $Q \in {\mathfrak B}(\Hi^{\otimes m})$ is} \emph{permutation invariant} if and only if it commutes with all the subsystem permutations, {\as i.e.~$U_{\pi} Q U_{\pi}^\dag = Q$ or $\rho = U_{\pi} \rho U_{\pi}^\dag$ for all $\pi \in {\mathfrak P}_m$}. Given any observable $Q\in \mathfrak{H}(\Hi^{\otimes m})$, we can obtain a permutation invariant observable $\bar{Q}$ by considering a projection of $Q$ onto the permutation-invariant set. By using basic group properties, one shows that \cite{mazzarella-TAC,mazzarella-symmetrization} the orthogonal projection with respect to the Hilbert-Schmidt inner product $\langle X_1, X_2 \rangle = \trace(X_1^\dag X_2)$ takes the form:
\begin{equation}
\label{permutinvL}
\bar{Q} = \bar\Ec(Q):=\frac{1}{m!} \, \sum_{\pi \in \mathfrak{P}_m} U_{\pi}\, Q\, U_{\pi}^\dag \, .
\end{equation}
This is a completely positive trace-preserving (CPTP) map.


\subsection{Continuous-time Quantum Dynamical Semigroups}
In this work we will consider finite-dimensional, time-homogenous
Markovian dynamics \cite{gorini-k-s,lindblad}. These correspond to continuous Quantum Dynamical Semigroups (QDSs) of CPTP maps, whose generator can always be expressed in the form ($\hbar \equiv 1$): \beq
\label{eq:lindblad}{\cal L}(\rho(t))
=-i[H,\,\rho(t)]+\sum_k \Big(
L_k\rho(t)L^\dag_k-\frac{1}{2}\{L_k^\dag L_k,\,\rho(t)\} \Big), \eeq
{\as with notations $[A,B] = AB-BA$ and $\{A,B\} = AB+BA$.} The QDS is thus obtained by formal exponentiation of ${\cal L},$ $\rho_t=e^{{\cal L}t}\rho,$ and ${\cal L}$ is uniquely determined by the Hamiltonian $H=H^\dagger$ and the set of \noise operators $\{L_k\}$. 

We are interested in the asymptotic behavior of QDSs in which the operators $H,\{L_k\}$ satisfy locality constraints.  Hence, we shall allow these semigroup generators to act in a non-trivial way only on certain predefined subsets of subsystems, which are called {\em neighborhoods} ${\cal N}_j$, $j=1,\ldots, M$ \cite{ticozzi-QL,ticozzi-QIP,mazzarella-TAC}. These can be specified as subsets of the set of indices labeling the subsystems:
\[{\cal N}_j\subseteq\{1,\ldots,n\}, \quad j=1,\ldots, M.\]
In analogy with the strictly local case and following \cite{ticozzi-QL}, we call a  \noise operator $L_k$ {\em Quasi-Local (QL)} if there exists a neighborhood ${\cal N}_j$ such that:
\[L_k= L_{{\cal N}_j}\otimes I_{\bar{\cal N}_j},\] 
where $L_{{\cal N}_j}$ accounts for the action of $L$ on the
subsystems included in ${\cal N}_j$, and $I_{\bar{\cal
N}_j}:=\bigotimes_{a\notin{\cal N}_j}I_a$ is the identity on the
remaining subsystems.  Similarly, a {\em Hamiltonian is QL} if it
admits a decomposition into a sum of QL terms: \[H=\sum_jH_j, \quad
H_j=H_{{\cal N}_j}\otimes I_{\bar{\cal N}_j}.\] 
\noindent 
A QDS generator \eqref{eq:lindblad} will be called QL if its Hamiltonian and all its \noise operators
are QL. The different definitions of QL for Hamiltonian and \noise operators is intrinsic to the way they enter the dynamics --- linearly for the former and as a convex sum of quadratic terms for the latter. While the decomposition into Hamiltonian and dissipative part of the generator $\mathcal{L}$ is not unique, the QL property is well defined since the freedom in the representation does not affect the tensor (locality) structure of $H$ and $\{L_k\}$.

This way of introducing locality constraints allows us to cover a number of specific locality definitions of interest in quantum condensed matter and quantum information applications, including nearest-neighbor interaction on a graph or lattice, and/or QL Hamiltonian components and noise generators that act non-identically on a number of no more than $t$ subsystems (so-called $t$-body interactions).



\section{DISTRIBUTED SYMMETRIZING DYNAMICS}

\subsection{The Task}

The problem we aim to address, which can be seen as a stabilization problem, is the following:

{\em Given a multipartite system as in Section \ref{sec:system} and a {\em fixed} locality notion associated to a neighborhood structure $\{{\cal N}_j\},$ find a QL generator ${\cal L}$ of the form \eqref{eq:lindblad} such that for every $\rho\in{\mathfrak D}(\Hi^{\otimes m})$ we have
\beq \label{eq:task}\lim_{t\rightarrow\infty} e^{{\cal L} t}\rho= \bar\Ec(\rho)=\frac{1}{m!} \, \sum_{\pi \in \mathfrak{P}} U_{\pi}^{\dagger}\rho U_{\pi} \, .
\eeq}



\subsection{Main Result: Algebraic Approach}\label{sec:main}


We shall obtain our symmetrizing QDS as a particular case of a more general construction. Consider a QDS generator
\beq\label{eq:lindbladG}{\cal L}_U(\rho(t))
=\sum_k \alpha_k \, \Big(
U_k\rho(t)U^\dag_k-\frac{1}{2}\{U_k^\dag U_k,\,\rho(t)\} \Big), \eeq
{\as with $\alpha_k>0$ for each $k$}; that is, a generator associated to a null Hamiltonian and a set of \noise operators that are unitary, $\{U_k\}\subset {\mathfrak U}(\Hi^{\otimes m}).$ This generator is in particular unital, {\em i.e.}~it preserves the identity. Consider ${\cal L}_U$ as acting on the whole ${\mathfrak B}(\Hi^{\otimes m})$, and not only on the density operators. The following holds. 

\begin{lemma}\label{lemma:commutant}
The set of fixed points of ${\cal L}_U$ corresponds to the commutant {\cal A'} of the algebra generated by the $\{ U_k\}$:
\beq\label{eq:fpset} {\cal A'}=\{X\in{\mathfrak B}(\Hi^{\otimes m})|X U_k = U_k X, \forall k\}.\eeq
\end{lemma}
\proof 
Notice that \eqref{eq:lindbladG}, thanks to the unitary character of the noise operators, boils down to the simple form:
\beq\label{Lsimpleform}{\cal L}_U(X)
=\sum_k \alpha_k (U_k X U^\dag_k - X)\,,\eeq
and hence ${\cal L}_U(X)=0$ if and only if $X$ is a fixed point of the unital CPTP map $\Ec_U(X)=\sum_k \alpha_k\, U_kXU^\dag_k.$ It has been shown in e.g.~\cite{viola-IPSlong,wolf-quantumchannels} that the set of invariant points for such unital CPTP maps is the commutant of the operators entering its representation. \hfill \qed

In order to construct an effective generator for our task with locality constraints, it is convenient to recall that the full permutation group is generated by the set of pairwise transpositions restricted to the edges of any spanning tree connecting the subsystems. More generally: if the neighborhood structure covers the whole network, and there does not exist a (nontrivial) partition of the neighborhoods into groups with non-overlapping supports, then it is easy to show that the set of {\em all the transpositions allowed by this locality notion} is sufficient to generate the whole permutation group. One proves this easily by noting that the corresponding set of all transpositions includes in particular a set of pairwise transpositions along a spanning tree. This observation motivates the construction of \noise operators implementing swaps of neighboring subsystem states.\\

{\em\as Consider ${\cal L}_U$ to be associated to allowed subsystem swap operations: denoting by $\mathfrak{P}_{\N_j} \subset \mathfrak{P}_m$ the set of all permutations that involve the integers labeling subsystems in $\N_j$ only, we take
\begin{equation}\label{eq:oursolution}
\{ U_k \} \subset \{ U_\pi \, \vert \, \exists j \text{ such that } \pi \in \mathfrak{P}_{\N_j} \} \, .
\end{equation}}

Building on the previous result, we can give the following characterization of invariant sets:
\begin{prop}
Consider a generator of the form \eqref{eq:lindbladG}, with $\{U_k\}$ chosen as in \eqref{eq:oursolution}. Then the set of fixed points for the associated dynamics \eqref{eq:lindbladG} is the set of permutation-invariant states if and only if $\{U_k\}$ generates the full permutation group.
{\as \newline Moreover, the dynamics keeps $\bar\Ec{X_t}=\bar\Ec{X_0}$ invariant for all $t$ and all $X_0 \in\mathfrak{B}(\Hi^{\otimes m})$.} \end{prop}
\proof
The `if' case (assuming that $\{U_k\}$ generate the full permutation group) is a direct consequence of Lemma \ref{lemma:commutant}, the set of fixed points is the set of permutation-invariant states. On the other hand, if the $\{ U_k \}$ generate a proper subgroup of the permutation group, then the invariant set under \eqref{eq:lindbladG} will, {\as according to Lemma \ref{lemma:commutant}, contain all states that are invariant under the subsystem permutations from that subgroup. By definition of a subgroup and of the action of permutations on $\mathfrak{B}(\Hi^{\otimes m})$,} this includes states that are not invariant under some $\pi \in \mathfrak{P}_m$.

By using basic group properties\footnote{\as In particular, for each selected permutation $\pi_1$ in ${\cal L}_U$, the sum in $\bar\Ec$ runs over all $\pi_1\,\pi$ with $\pi \in \mathfrak{P}_m$, which is just the same as over all $\pi \in \mathfrak{P}_m$.} it is easy to see that:
\[\bar\Ec({\cal L}_U(X))={\cal L}_U(\bar\Ec(X)) \]
for all $X \in \mathfrak{B}(\Hi^{\otimes m})$. This readily implies
$$ \bar\Ec(e^{{\cal L}_U t}\rho) \, = \, e^{{\cal L}_U t}\bar\Ec(\rho) \, =\, \bar\Ec(\rho) \, ,$$
where in the last equality we used the fact that $\bar\Ec(\cdot)$ projects on the set of fixed points. \hfill \qed

In the case where the $U_k$ only comprise pairwise swaps, the transposition matrices $U_k$ are Hermitian, and the same holds for the (super-)operator ${\cal L}_U.$ {\as If moreover the $\alpha_k$ are time-invariant, then the real spectrum of ${\cal L}_U$ allows us to conclude that the set of fixed points coincides with the globally asymptotic stable set~\cite{baumgartner-2}, and we conclude that \eqref{eq:task} indeed holds. However, a more general result can be established to highlight the robustness of quantum symmetrization.}



{\as \subsection{Extension via a Lyapunov Approach}\label{sec:main2}

We want to establish the convergence property of the QDS generator \eqref{eq:lindbladG},\eqref{eq:oursolution} under general conditions that remind classical consensus: the $\alpha_k$ (and the allowed neighborhoods) can be time-varying, and the local permutation operators that are included may involve more than pairwise swaps. The latter setting is probably less attractive in practice, as pairwise physical interactions would be sufficient and probably the easiest to implement. {\ft However, this would indeed give rise to the possibility of implementing any \emph{balanced} graph in the quantum setting,} and not just \emph{undirected} graphs. 

For balanced graphs, it is known in classical consensus theory \cite{ConsensusReview} that several Lyapunov functions can be used to prove convergence independently of the  $\alpha_k$.
{\ft  In this section, we propose two Lyapunov functions to show that indeed \eqref{eq:lindbladG},\eqref{eq:oursolution} achieve our goal \eqref{eq:task} under similar conditions{\ft, and  much more: {\em any dynamics of that form converges to the set of operators that are invariant with respect to the group generated by the unitary Lindblad operators.} The first one is the Hilbert-Schmidt distance between $\rho_t$ and its symmetrized {image}, while the second one uses a relative entropy distance between the propagator and the  symmetrizing {projector} via a suitable lift of the dynamics.} 
 
\begin{prop}\label{propA}
The Hilbert-Schmidt distance to consensus $V(\rho) = \tfrac{1}{2} \trace\left((\rho-\bar\Ec(\rho))^2 \right)$ is a strict Lyapunov function for \eqref{eq:lindbladG} with respect to the set of its fixed points, for any neighborhood specifications and $\alpha_k > 0$.
\end{prop}}
\proof
We have
\begin{eqnarray*}
\frac{dV}{dt}(\rho) & = & \trace\left((\rho-\bar\Ec(\rho))(\tfrac{d\rho}{dt}-\bar\Ec(\tfrac{d\rho}{dt})) \right)\\
& = & \sum_k \alpha_k \, \trace\left((\rho-\bar\Ec(\rho))\phantom{(U_k \rho U_k^\dag-\rho-\bar\Ec(U_k \rho U_k^\dag)+\bar\Ec(\rho))}\right.\\
& & \phantom{\sum_k \alpha_k \, \trace}\;\;\;\left. \vphantom{(\rho-\bar\Ec(\rho))}(U_k \rho U_k^\dag-\rho-\bar\Ec(U_k \rho U_k^\dag)+\bar\Ec(\rho)) \right)\\
& = & - \sum_k \alpha_k \, \trace\left(\rho(\rho-U_k\rho U_k^\dag)\right)\\
& = & - \sum_k \alpha_k \, \trace\left( (\rho-U_k\rho U_k^\dag)^2\,/2 \right)\, .
\end{eqnarray*} 
The first equality is obtained by linearity; the third equality uses that, from the observation of footnote 1, it follows that $\bar\Ec(U_k \rho U_k^\dag) = \bar\Ec(\rho)$ when $U_k$ is a permutation operator, and that $\trace(\bar\Ec(\rho) U_k \rho U_k^\dag) = \trace(\bar\Ec(\rho) \rho)$ since $U_k^\dag \bar\Ec(\rho) U_k = \bar\Ec(\rho)$. The last equality uses standard trace and unitary operator properties. {\ft From the last expression, it is clear that $V$ is monotone non-increasing, and that $dV/dt=0$ only when $\rho$ belongs to the set of symmetric states.} \hfill \qed

While Proposition \ref{propA} is sufficient to characterize convergence of our quantum consensus algorithm, we can in fact present a stronger version of it {\ft that ensures convergence with respect to any action of the same group,} see \cite{mazzarella-symmetrization,mazzarella-CDC}. {\ft This is done by lifting the dynamics \eqref{eq:lindbladG},\eqref{eq:oursolution} to convex weights on the set of permutations of $m$ integers. }
{\ft The first step is to show that we can restrict our attention to dynamics that are convex combinations of subsystem permutations or, more in general, convex combinations of the group generated by the unitaries in \eqref{eq:lindbladG}.}

{\ft \begin{prop}\label{redsym} For all $t,$ there exists a vector $\pope \in \mathbb{R}^{m!}$, whose elements are indexed by all the permutations $\pi \in \mathfrak{P}_m$ and with $\pope_\pi \geq 0$ $\forall \pi$, $\;\sum_\pi \, \pope_\pi = 1$, such that
\beq\label{osrform}\rho_t = \sum_{\pi \in \mathfrak{P}_m} \, \pope_\pi(t) \, U_\pi \rho_0 U_\pi^\dag \, .\eeq \end{prop}
\proof {For $t=0$ it is obvious, with $\pope_{\rm e}=1$ and the other elements equal to zero. 
If we now apply ${\cal L}_U$ to a state of the form $\eqref{osrform},$ by using basic group properties we obtain:
\[{\cal L}_U=\sum _{\pi \in \mathfrak{P}_m} \, \tfrac{d}{dt} \pope_\pi(t) \, U_\pi \rho_0 U_\pi^\dag,\]
where
\[\tfrac{d}{dt} \pope_\pi(t) = \sum_{k} \, \alpha_k \; (\pope_{k^{-1} \pi} - \pope_{\pi}). \;\]
It is then easy to see, {\em e.g.} by considering the propagator as the exponential of ${\cal L}_U$, that the corresponding dynamics is still of the form \eqref{osrform}. In addition, given the form of the derivative, it is immediate to see that the minimal weight $\pope_\pi$ can only increase. Since the initial minimal weight is zero, the weights remain positive at all times. Lastly, their sum must remain equal to one in order for the map to be trace preserving. 
\qed}

In the proof of Proposition \ref{redsym}, we show that the dynamics \eqref{eq:lindbladG},\eqref{eq:oursolution} is equivalent to the dynamics of $\pope(t)$ following:
\begin{equation}\label{eq:pope-dyns}
\tfrac{d}{dt} \pope_\pi = \sum_{k} \, \alpha_k \; (\pope_{k^{-1} \pi} - \pope_{\pi}) \; ,
\end{equation}
\noindent where the composition $k^{-1} \pi$ must be understood in the group-inverse and group-multiplication sense. Note that this lift is not unique. The advantage of this viewpoint (beyond the fact that \eqref{eq:pope-dyns} looks more like a standard consensus algorithm) is that \eqref{eq:pope-dyns} can be analyzed \emph{independently of the group action} of $\pi$ --- be it through conjugate action of $U_\pi$ on $\rho$ like here, or acting on classical states or probability distributions or any other interesting space. In fact it is even not essential that $\pope$ is indexed by permutations, any discrete group could be used, as we highlight in \cite{mazzarella-symmetrization}.

Our purpose now is to establish the convergence of the lift \eqref{eq:pope-dyns} with a Lyapunov function, showing that the conclusions of \cite{mazzarella-symmetrization} hold also in continuous-time.\\

\begin{prop}\label{propB}
The Kullback-Leibler divergence of $\pope$ with respect to the uniform vector $D(\pope) = \sum_{\pi \in \mathfrak{P}_m} \, \pope_\pi (\,\log \pope_\pi - \log(\tfrac{1}{m!})\,)$ is a strict Lyapunov function establishing convergence of \eqref{eq:pope-dyns} towards $\pope_\pi=1/{m!}$, for any neighborhood specifications and $\alpha_k > 0$.
\end{prop}
\proof
We have
\begin{eqnarray*}
\frac{d}{dt}D(\pope) & = & \sum_{\pi \in \mathfrak{P}_m} \, \tfrac{d\pope_\pi}{dt} (1 + \log \pope_\pi - \log(\tfrac{1}{m!}))\\
& = & \sum_{\pi \in \mathfrak{P}_m,\; k} \alpha_k \, (\pope_{k^{-1}\pi}-\pope_\pi)(1 + \log \pope_\pi - \log(\tfrac{1}{m!}))\\
& = & \sum_{\pi \in \mathfrak{P}_m,\; k} \alpha_k \, (\pope_{k^{-1}\pi}-\pope_\pi) \log \pope_\pi \\
& = & -\sum_{\pi \in \mathfrak{P}_m,\; k} \alpha_k \, \pope_\pi (\log \pope_\pi-\log \pope_{k\,\pi})\\
& = & -\sum_{k} \alpha_k \, K(\pope \Vert \Pi_k(\pope)), \, 
\end{eqnarray*}
{\ft where $\Pi_k$ is the linear operator that maps the value of $\pi$-th component  of $\pope$ to component $k\,\pi$, for all $\pi$, and $K(\pope^A \Vert \pope^B)$ denotes the relative entropy $\sum_{\pi \in \mathfrak{P}_m} \, \pope^A_\pi (\,\log \pope^A_\pi - \log(\pope^B_\pi)\,)$. The third equality is obtained by observing that $\sum_{\pi \in \mathfrak{P}_m} \pope_\pi = \sum_{\pi \in \mathfrak{P}_m} \pope_{k^{-1} \pi}=1$ for each $k$, such that the multiples of the terms $1$ and $-\log(\tfrac{1}{m!})$ cancel out. The fourth equality implements the change of variable $\pi \rightarrow k^{-1} \pi$, noting that this does not require to change the argument of the sum over $\mathfrak{P}_m$. }

Since the relative entropy is known to be non-negative and equal to zero if and only if $\pope^A=\pope^B$, $D$ is monotone non-increasing and will stop decreasing only when the vector $\pope$ is invariant under the action of $\Pi_k$ for all $k$ in our actions set. In our particular case, where $k$ are transpositions, this implies by standard Lyapunov argument that we reach the unique fixed point of the lifted dynamics. \qed

The last result is given in terms of the lifted dynamics: when the unitaries used in ${\cal L}_U$ with non-zero weights  generate the whole subsystem-permutation group for the original dynamics, it directly implies that the propagator converges to the map \eqref{permutinvL}.

{\ft We thus proved convergence to symmetric states under quite general setting, that could be further extended by considering {\em time-varying} dynamics. The results on the Lyapunov functions hold with time-varying, non-zero $\alpha.$ Combining these with standard limit-set arguments from consensus \cite{ConsensusReview}, it can be shown that {\em e.g.}~in the case of pairwise interactions associated to an undirected graph (where subsystems are nodes and $\alpha_\pi$ can be different from zero only if there is a link between the subsystem swapped by $\pi$), our dynamics would drive any $\rho_0$ towards $\bar\Ec(\rho_0)$, i.e.~\emph{\eqref{eq:task} is satisfied, if there exist finite $T,\underline{\alpha}>0$ such that the edges for which $\int_t^{t+T} \alpha(\tau) d\tau > \underline{\alpha}$ form a connected graph.}}



\section{APPLICATIONS}

{\as {\ft We next employ the proposed symmetrizing dynamics} to two applications, already presented in \cite{mazzarella-TAC} in the discrete-time context. The latter called for adding algorithmic steps around the symmetrization procedure. In contrast, the continuous-time operation allows to propose one integrated implementation, where application-specific elements act simultaneously, like perturbing dynamics, with the symmetrization.}


\subsection{Pure State Preparation with a ``Stubborn'' Subsystem}
Design techniques for the asymptotic preparation of any pure state in a {\em single system} \cite{bolognani-arxiv} using QDS are well-known \cite{ticozzi-QDS,ticozzi-markovian,ticozzi-generalstate}. We here show how to polarize the population of a quantum network towards a target pure state by {\em locally} perturbing our symmetrizing dynamics.

Consider again the system of Section \ref{sec:system} and the generator we constructed in Section \ref{sec:main}. We are interested in stabilizing the whole quantum network into a pure state. It can be shown that {\em any pure, factorized state of the form:}
\[ \hat\rho=\ket{\psi}\bra{\psi}\otimes\cdots\otimes\ket{\psi}\bra{\psi},\] 
with given $\q{\psi}$, can be asymptotically obtained by adding a single, strictly local \noise operator to ${\cal L}_U.$
This operator acts on a single system, and has the target local pure state as its unique invariant state:
\begin{equation}\label{eq:Ltot1}
{\cal L}_{\rm tot}={\cal L}_U+{\cal L}^{(j)}\otimes {\cal I}_{\bar j},
\end{equation}
with ${\cal L}^{(j)}$ such that for all $\rho\in{\mathfrak D}(\Hi_j),$ we have
\[\lim_{t\rightarrow\infty}e^{{\cal L}^{(j)}t}\rho=\ket{\psi}\bra{\psi}.\] Simple ways on how to construct such a generator, with or without feedback, have been presented in \cite{ticozzi-QDS,ticozzi-markovian,ticozzi-generalstate,ticozzi-NV}. {\as Thus $j$ acts as ``stubborn'' subsystem that is continuously attracted towards $\q{\psi}\qd{\psi}$.}

The two terms in \eqref{eq:Ltot1} implement two partially competing dynamics, one associated to ${\cal L}_U$ that drives every state into the symmetric set, and another one associated to ${\cal L}^{(j)}$ that drives every state towards something of the form $\ket{\psi}\bra{\psi}\otimes\tau_{\bar j},$ without affecting $\tau_{\bar j} \in {\mathfrak D}(\Hi^{\otimes(m-1)}).$ {\as Since those two dynamics are linearly independent whenever they are nonzero,} the only invariant set for the combined dynamics is the intersection of the two limit sets, that is the target state $\hat\rho.$ A standard Lyapunov argument and LaSalle invariance theorem (see e.g.~\cite{khalil-nonlinear}), just considering the trace distance with respect to $\hat\rho$, allows to conclude that $\hat\rho$ is indeed asymptotically prepared by the QDS associated to ${\cal L}_{\rm tot}$.

By variations of this method, the same control capabilities can be used to engineer dynamics that asymptotically drive the state of the quantum network to have support on an arbitrary target subspace of the network's joint Hilbert space, provided it is invariant with respect to subsystem permutations.


\subsection{Estimation of the Network Size}

Consider again a set of $m$ identical subsystems as in Section \ref{sec:system} and the ability of turning on and off the symmetrizing generator ${\cal L}_U$. In addition, assume that we can only access the first $p$ subsystems: on these subsystems we can implement an action of local ${\cal L}^{(j)}$ stabilizing the single systems into desired pure states, as in the previous application, {\ft as well as} measurements of identical, non-degenerate, purely local observables $Q \in \mathfrak{H}(\Hi)$.
While we assume to know the number $p$ of \emph{accessible} subsystems, we are now interested in estimating the \emph{unknown total number} $m$ of subsystems in the quantum network.

{\as To do so in one run without access to anything else than the $p$ first subsystems, we need to know something about the initial state of the whole network: we will assume that \emph{all subsystems are initially prepared in a state that has orthogonal support to a fixed ``marker'' eigenstate $\q{\psi}$ of $Q$}. If for some reason this situation cannot be naturally assumed, then it can be asymptotically obtained with the protocol described in the previous section\footnote{It is worth noting that, while asymptotic exact preparation of a state or a subspace would ideally entail an infinite evolution time, convergence to the stable set is exponential and hence errors can be made arbitrarily small in finite time.} and a target state $\ket{\phi}\bra{\phi}$ with $\langle \phi \vert \psi \rangle = 0$.

In order to estimate the size of the sample, we can then implement the following dynamics:}

\begin{description}
\item {\em 1 - Preparation:} first prepare the network in a state $\rho'=\bigotimes_j\q{\phi}\qd{\phi})^{(j)},$ that has orthogonal support to the  ``marker'' eigenstate $\q{\psi}$ of $Q$, $\braket{\psi}{\phi}=0$. 
\item{\em 2 - Perturbation:} Next, reset each of the $p$ probe subsystems {\em in the marker eigenstate $\q{\psi}$ of $Q$};

\item{\em 3 - Symmetrization} Let the network evolve with ${\cal L}_U$ to the symmetric set;

\item{\em 4 - Readout} Perform measurements of $Q$ on the $p$ probe subsystems, recording how many times $\q{\psi}$ is obtained.
\end{description}

\noindent The first two steps prepare the network into a state
$$\q{\psi}\qd{\psi} \, \otimes \, ... \, \otimes \, \q{\psi}\qd{\psi}\; \otimes \; \q{\phi}\qd{\phi} \, \otimes \, ... \, \otimes \, \q{\phi}\qd{\phi}\ \; .$$
The statistics of measuring $Q$ on the $p$ probe subsystems after Step 3, equals the statistics of measuring $Q$ before Step 2 on $p$ uniformly randomly selected subsystems. In the latter case, whenever one of the first $p$ subsystems was selected we would get outcome $\q{\psi}$, while whenever a subsystem $j>p$ is selected we would certainly not get $\q{\psi}$. The random variable $K$ counting the number $k$ of times $\q{\psi}$ is detected in Step 4 therefore follows a hypergeometric distribution,
$$K=k \text{ with probability } (^p_k )\, (^{m-p}_{p-k} ) \; / \; (^m_p )$$
where $(^b_a) = b!/(a!(b-a)!)$. 
{\ft Following \cite{mazzarella-TAC}, we can show that

\[\mathbb{E}[K]=p^2 / m,\]

\noindent and hence the candidate estimator for $m$ can be chosen to be $\hat m=p^2/\hat K,$ where $\hat K$ is the sampled value of $K.$ It is then easier to study the statistical properties of $\hat m^{-1},$ being just a rescaling of the measured $\hat{K}.$ It is then possible to show that the {\em relative error} $\frac{\hat m^{-1}- m^{-1}}{ m^{-1}}$ of $\hat{m}^{-1}$ has mean zero, i.e. it is an unbiased estimator. Its variance can be computed and has the form:}
\begin{eqnarray}\label{varr}
\mathbb{E}\left[\left(\frac{\hat m^{-1}- m^{-1}}{ m^{-1}}\right)^2\right]
&=& \frac{(m-p)^2}{p^2(m-1)}.
\end{eqnarray}}

{\ftl This shows that if we consider $p=\alpha\cdot m$ to be a {\em fixed} fraction of the total population, when the population increases the variance \eqref{varr} goes to zero as $1/m$. Then for the limit of large $m$, we can conclude that the variance of $\hat m$ also goes to zero as $1/m$.}

%


%

\section*{ACKNOWLEDGMENT}

The authors whish to thank Prof. Lorenza Viola for her collaboration on the study of quasi-local Lindblad dynamics, and for stimulating discussion on the topic of symmetrization and consensus for quantum networks.


\end{document}